\documentclass[12pt]{article}
\usepackage{epsfig}
\usepackage{amsfonts}

\def\be{\begin{equation}}
\def\ee{\end{equation}}
\def\ba{\begin{eqnarray}}
\def\ea{\end{eqnarray}}

\def\nn{\nonumber}

\def\sst{\scriptscriptstyle}

\def\R{{\cal R}}
\def\F{{\cal F}}

\def\M{{\cal M}}

\def\K{{\cal K}}
\def\N{{\cal N}}

\def\Mbar{\bar{\M}}

\def\Nbar{\bar{\N}}

\def\mubar{\bar{\mu}}
\def\nubar{\bar{\nu}}
\def\Atot{{\cal A}_{\rm tot}}
\def\bigL{{\bf L}}

\def\wbar{\bar w}
\def\abar{\bar a}

\def\Foneinst{{\cal F}_1}

\def\1I{\mbox{{\tiny 1-I}}}

\def\hyp{{\rm hyp}}
\def\Lambdahyp{{\Lambda_\hyp}}

\def\bigL{{\bf L}}
\def\Lambdabar{\bar\Lambda}

\def\dalpha{{\dot\alpha}}
\def\dbeta{{\dot\beta}}

\def\sigmabar{\bar\sigma}

\def\D{{\cal D}}
\def\Dslash{\,\,{\raise.15ex\hbox{/}\mkern-12mu \D}}
\def\Dbarslash{\,\,{\raise.15ex\hbox{/}\mkern-12mu {\bar\D}}}

\def\hf{{\textstyle{1\over2}}}
\def\sqrtwo{\sqrt{2}\,}

\def\Im{{\rm Im}}
\def\Re{{\rm Re\,}}

\def\v{{\rm v}}
\def\vbar{\bar{\rm v}}
\def\zbar{\bar z}
\def\alphabar{\bar\alpha}

\newcommand{\rf}[1]{(\ref{#1})}

\begin{document}
\baselineskip 6.8mm

\begin{titlepage}
\begin{flushright}
hep-th/9912011\\
December 1999\\
\end{flushright}
\begin{centering}
\vspace{.4in}
{\Large {\bf One-instanton test of the exact prepotential for
$\N=2$\\
SQCD coupled to a symmetric tensor hypermultiplet}}\footnote{Work
supported by the European Commission TMR programme ERBFMRX-CT96-0045.}\\
\vspace{.5in}
{\bf Matthew J. Slater} \\
\vspace{.1in}
INFN - Laboratori Nazionali di Frascati\\
P.O. Box 13, I-00044 Frascati (Roma), Italy \\
\texttt{slater@lnf.infn.it}\\
\vspace{.5in}
{\bf Abstract}\\
\end{centering}
Using the ADHM instanton calculus, we evaluate the one-instanton contribution
to the low-energy effective prepotential of $\N=2$ supersymmetric $SU(N)$
Yang-Mills theory with $N_F$ flavors of hypermultiplets in the fundamental
representation and a hypermultiplet in the symmetric rank two
tensor representation. For $N_F<N-2$, when the theory is asymptotically free,
our result is compared with the
exact solution that was obtained using \mbox{M-theory} and we find complete
agreement.

\end{titlepage}

An important development in the study of four-dimensional
Yang-Mills theory has been the prediction of exact results for models
with $\N=2$ supersymmetry.
The pioneering work was that of Seiberg and Witten,
who investigated the pure $SU(2)$ Yang-Mills theory~\cite{SW1}
and also $SU(2)$ models with matter transforming in the
fundamental or the adjoint representation~\cite{SW2}.
Subsequently, their
results have been generalized to a wide range of models with
simple or direct product gauge group and matter in various
representations.

The general prescription for the exact results involves
an algebraic curve, whose form depends on the model
being considered, and a particular one-form defined on the curve.
The one-form is to be integrated over one-cycles that form a
canonical basis on the curve. This yields an exact solution
for the effective prepotential, the holomorphic function that
describes the low-energy Coulomb branch physics.

A general property of the effective prepotential is that
at weak-coupling it has an expansion consisting of a one-loop
perturbative term plus an infinite series of non-perturbative
terms that correspond to instanton effects. In principle, it is possible to
directly calculate these effects from first principles
in the weakly-coupled quantum field theory. This is an important
exercise, since by comparing the results of first-principles
instanton calculations with the predictions extracted from the curves
one can directly test the exact results and the means 
by which they have been derived. Over the last few years there has
arisen a program of research concerned with carrying out precisely such
tests of the exact results~[3--10].

In this Letter, we continue this program of instanton tests.
We focus on a model which has gauge group $SU(N)$ and
$N_F$ flavors of hypermultiplets in the fundamental
representation as well as a hypermultiplet
transforming in the symmetric rank two tensor representation.
In other words, we investigate $\N=2$ supersymmetric QCD ($\N=2$ SQCD)
coupled to a symmetric tensor hypermultiplet. 
The algebraic curve corresponding to this model was derived by
Landsteiner, Lopez and Lowe by studying brane configurations in
M-theory~\cite{LLL}.
The one-instanton contribution to the effective prepotential
predicted by the curve has been explicitly determined by
Ennes et al.~\cite{ENRS}. 
To test the curve, we calculate the same contribution from
first principles.
Our result completely agrees with the curve prediction
thus verifying the exact solution and the M-theory method at the
one-instanton level.\footnote{Our attention is restricted to
$N_F<N-2$, when the theory is asymptotically free; we are unable
to provide a test in the vanishing $\beta$-function case $N_F=N-2$
since here the curve of~\cite{LLL} has not been explicitly parameterized.
Instanton tests of the exact results for $\N=2$ SQCD have revealed
important discrepancies in the
vanishing $\beta$-function case that --- at least for $N>3$ ---
have yet to be completely resolved~\cite{dkm34,dkmfive,KMS}.}

To perform the instanton calculation, we employ the instanton calculus,
based on the construction of Atiyah, Drinfeld, Hitchin and
Manin (ADHM)~\cite{ADHM}, that has been developed
in~\cite{dkmone,dkm34,dkmten,KMS}. Whilst this calculus
is important mainly for the study of {\it multi}-instanton
effects,
it is useful in the present context
because it greatly facilitates the collective coordinate integration
at the heart of the instanton calculation.
(In particular, it simplifies an otherwise highly
non-trivial integration over the group space collective coordinates of
the instanton.)

To begin, we will briefly summarize the essential
features of the ADHM instanton
calculus for the specific case of $\N=2$ SQCD~\cite{dkm34,dkmten,KMS},
borrowing from the results of Ref.~\cite{KMS} in particular.
We will subsequently describe how this calculus is extended
when $\N=2$ SQCD is coupled to a symmetric tensor
hypermultiplet. We specialize
to the one-instanton case throughout, this being sufficient for our purpose.

In $\N=2$ SQCD, the supersymmetric ADHM one-instanton background
is parameterized by:
i) an $(N+2)\times 2$ complex-valued matrix $a$
(and its conjugate $\abar$), associated with the gauge field,
ii) a pair of $(N+2)$-dimensional Grassmann-valued vectors
$\{\M,\N\}$ (and the conjugate pair $\{\Mbar,\Nbar\}$),
associated with the two adjoint Weyl fermions, and iii) $N_F$
pairs of Grassmann parameters $\{\K_f,\tilde{\K}_f\}$ ($f=1,\ldots,N_F$),
associated with the
$N_F$ flavors of fundamental and conjugate-fundamental Weyl fermions.
Whereas $\K_f$ and $\tilde{\K}_f$ are free parameters, the
collective coordinates associated with the adjoint fields,
$a$, $\M$ and $\N$, are required to satisfy certain constraint equations.
To write these equations, it is first useful to decompose $a$, $\M$ and $\N$
(and their conjugates) as follows:
\ba
a_{\lambda\dalpha} = \pmatrix{ w_{u\dalpha} \cr \cr
{a'}_{\alpha\dalpha} },\qquad && \qquad
\abar^{\dalpha}_{\ \lambda} = \left( \wbar^{\dalpha}_{\ u} \ \
\abar^{\prime\,\dalpha\alpha} \right),\nn\\
\M_{\lambda} = \pmatrix{ \mu_{u} \cr \cr
{\M'}_{\alpha} },\qquad && \qquad
\Mbar_{\lambda} = \left( \mubar_{u} \ \
\Mbar^{\prime\,\alpha} \right),\nn\\
\N_{\lambda} = \pmatrix{ \nu_{u} \cr \cr
{\N'}_{\alpha} },\qquad && \qquad
\Nbar_{\lambda} = \left( \nubar_{u} \ \
\Nbar^{\prime\,\alpha} \right),
\ea
where the ranges of the various indices are
$1\leq\lambda\leq N+2$, $1\leq u\leq N$ and $\alpha,\dalpha=1,2$.
Correspondingly, the constraints on $a$ read
\ba
{a'}_{m}&=&{\abar'}_{m},\\
\label{spin1const}
(\tau^c)^{\dalpha}_{\ \dbeta}\,
\wbar^{\dbeta}_{\ u} w_{u \dalpha} &=& 0,
\ea
where the $\tau^c$ ($c=1,2,3$) are the Pauli matrices and
$a^{\prime}_m$ and $\bar{a}^{\prime}_m$
are defined by the quaternionic expansions
$a^{\prime}_{\alpha \dalpha}=a^{\prime}_m \sigma^{m}_{\alpha \dalpha}$
and $\abar^{\prime\, \dalpha \alpha}=\abar^{\prime}_{m}
\sigmabar^{m\, \dalpha \alpha}$.
The constraints on $\M$ and $\N$ read
\ba
&&\ \ {\M'}_{\alpha}={\Mbar'}_{\alpha},\qquad\qquad\qquad\ \
{\N'}_{\alpha}={\Nbar'}_{\alpha},\\
\label{spin1/2const}
&&\mubar_{u}w_{u\dalpha} = -\wbar_{\dalpha u}\mu_{u},\qquad\qquad
\nubar_{u}w_{u\dalpha} = -\wbar_{\dalpha u}\nu_{u}.
\ea

Besides these collective coordinates, the ADHM one-instanton background in
$\N=2$ SQCD is further described by a parameter $\Atot$ associated with the
adjoint scalar field~$A$.
On the Coulomb branch, this field satisfies
\be
\langle A\rangle = \mbox{diag}(\v_1,\v_2,\ldots,\v_N),
\ee
where the $\v_u$ are arbitrary complex numbers that sum to zero.
The parameter $\Atot$ is determined in terms of these vacuum expectation
values (VEVs) and the collective coordinates $a$, $\M$ and $\N$ by the
following equation:
\be
\label{spin0const}
\bigL\cdot\Atot=\Lambda+\Lambda_f,
\ee
where $\bigL=\wbar^{\dalpha}_{\ u} w_{u\dalpha}$ and
\ba
\Lambda&=&i\wbar^{\dalpha}_{\ u} \v_u w_{u\dalpha},\\
\Lambda_f&=&\frac{1}{2\sqrt{2}}
\left(\mubar_u \nu_u-\nubar_u \mu_u \right).
\ea

The ADHM instanton calculus provides expressions for both the action 
and the collective coordinate integration measure corresponding to the
supersymmetric ADHM instanton background.
In terms of the above quantities,
the one-instanton action for $\N=2$ SQCD is given by~\cite{dkm34,KMS}
\ba
S_{\sst SQCD}^{\1I} &=& {8 \pi^2 \over g^2} +  
8\pi^2 |\v_u|^2 \wbar^\dalpha_{\ u} w_{u\dalpha} 
- 2\sqrtwo \pi^2 i\left( \mubar_{u} \vbar_u \nu_{u}
- \nubar_{u} \vbar_u \mu_{u}  \right)\nn\\
&& - 8\pi^2(\Lambdabar+\Lambdahyp)\Atot + \pi^{2}\sum_{f=1}^{N_{F}}m_{f}
\tilde{\cal K}_{f}{\cal K}_{f},\label{SQCDact}
\ea
where the $m_f$ are the bare masses of the hypermultiplets and
\be
\Lambda_{\hyp}=\frac{i}{4\sqrt{2}}\sum_{f=1}^{N_F}\K_f\tilde{\K}_f.
\ee
This fermion bilinear is associated with the conjugate
adjoint scalar field $A^\dagger$ in the same way that 
$\Lambda_{f}$ is associated with $A$; it reflects a Yukawa
source term involving the fundamental fermions in the
Euler-Lagrange equation for this field~\cite{dkm34,KMS}.

The one-instanton collective coordinate integration measure
for $\N=2$ SQCD takes the form of a flat measure with respect to the parameters
$\{a,\M,\N,\Atot,\K,\tilde{\K}\}$, with the constraint
equations~\rf{spin1const},~\rf{spin1/2const} and~\rf{spin0const}
imposed by means of $\delta$-functions under the integral
sign~\cite{dkmten,KMS}:
\ba
\int d\mu_{\sst SQCD}^{\1I} &=& \frac{C_1^{\prime}}{2\pi}
\frac{1}{\pi^{2N_F}}
\int d\Atot \,d^{2N}w \,d^{2N}\wbar\, d^{N}\mu\, d^{N}\mubar\,
d^{N}\nu\, d^{N}\nubar\,d^4a' \,d^2\M' \,d^2\N' \,d^{N_F}\K\,
d^{N_F}\tilde{\K} \nn\\
&& \times \, \, \delta\left(\bigL\cdot\Atot-(\Lambda+\Lambda_f)\right)
\left[ \prod_{c=1}^{3} \delta\left(\hf(\tau^c)^\dalpha_{\ \dbeta}\,
\wbar_{\ u}^\dbeta w_{u\dalpha}\right)\right]\nn\\
&& \times \left[ \prod_{\dalpha=1,2}
\delta\left(\mubar_u w_{u\dalpha}+\wbar_{\dalpha u}
\mu_u\right) \delta\left(\nubar_u w_{u\dalpha}+\wbar_{\dalpha u}
\nu_u\right)\right].\label{SQCDmeas}
\ea
The constant prefactor $C^{\prime}_1$ can be determined by
comparing with the standard 't Hooft-Bernard one-instanton
measure~\cite{tHooft,Bernard}. 
Thus, in the Pauli-Villars regularization scheme, one finds
$C^{\prime}_1=2^2\pi^{-2N}M^{2N-N_F}$, where $M$ is the regularization
mass.

Having summarized the ADHM instanton calculus for $\N=2$ SQCD, we
now describe its extension when this theory is coupled to a
symmetric tensor hypermultiplet.
The supersymmetric instanton background for the resulting
theory will include configurations corresponding to the component
fields of the extra hypermultiplet.
The relevant fermionic field configurations are given by the ``zero-mode''
solutions to the massless Dirac equation in the symmetric tensor
or conjugate-symmetric tensor representation. 
At the one-instanton level, these solutions are parameterized by a pair of
$(N+2)$-dimensional Grassmann-valued vectors, $\R$ and $\tilde{\R}$~\cite{CGT};
these can be decomposed as
\be
\R_{\lambda}  = 
\pmatrix{ \rho_{u} \cr \cr \R^{\prime}_\alpha },\qquad\qquad
\tilde{\R}_\lambda =
\left( \tilde{\rho}_{u} \ \ {\tilde{\R}}^{\prime\alpha} \right).
\ee
A useful simplification is that there are no constraints on
these collective coordinates~\cite{CGT}; the number of parameters in each of
$\R$ and $\tilde{\R}$ precisely equals the
number of zero-modes of the Dirac operator in the
symmetric tensor representation.\footnote{This
is true {\it only} at the one-instanton level; at higher order instanton
levels the collective coordinate matrices that parameterize the symmetric
tensor fermion zero-mode solutions  must indeed satisfy constraint
equations~\cite{CGT}.}

We can directly associate a collective coordinate integration measure
with the extra fermion zero-modes.
Since $\R$ and $\tilde{\R}$ are free parameters,
it is simply the (suitably normalized) flat measure
\be\label{hypmeas}
\int d\mu^{\1I}_{\hyp} = \frac{M^{-(N+2)}}{2^{2N}\pi^{2(N+2)}}
\int d^{N} \rho \, d^{N}\tilde{\rho} \, d^{2}\R^{\prime}\,
d^{2}\tilde{\R}^{\prime}.
\ee
The full one-instanton integration measure for the theory is
now given by the product of this and the full $\N=2$ SQCD
measure~\rf{SQCDmeas}.

The one-instanton action for the theory is straightforward to determine
using the methods of~\cite{dkmone,dkm34,KMS}. It is the sum of
the $\N=2$ SQCD action~\rf{SQCDact} and the hypermultiplet-induced contribution
\be
\Delta S_{\hyp}^{\1I}=-4\sqrt{2}\pi^2\tilde{\rho}_u
\v_u\rho_u-8\pi^2\Lambda_{\hyp}^{\prime}\Atot
+2m\pi^2(2\tilde{\rho}_u\rho_u
+\tilde{\R}^{\prime\alpha}\R^{\prime}_{\alpha}),
\ee
where $m$ is the bare mass of the symmetric tensor hypermultiplet and
\be
\Lambda^{\prime}_{\hyp} = -\frac{i}{\sqrt{2}}\tilde{\R}_{\lambda}{\R}_\lambda.
\ee
The origin of this fermion bilinear is similar to that of $\Lambda_{\hyp}$;
it reflects a Yukawa source term involving the symmetric tensor fermions
that appears in the Euler-Lagrange equation for $A^\dagger$.

In terms of the instanton action and collective coordinate integration
measure, one can derive the following expression
for the one-instanton contribution to the effective prepotential
of the theory~\cite{dkmtwo}:
\be
\label{F1intexp}
\F_1  = 8\pi i\int d{\tilde{\mu}}_{\sst SQCD}^{\1I} \,
d\mu_{\hyp}^{\1I} \,
\exp(-(S^{\1I}_{\sst SQCD} + \Delta S_{\hyp}^{\1I})).
\ee
Here $\int d{\tilde{\mu}}_{\sst SQCD}^{\1I}$ is a reduced version of
the full $\N=2$ SQCD measure,
which excludes the integration over the global position
of the instanton in $\N=2$ superspace, as represented
by $(x_0,\xi_1,\xi_2)=(a',\M'/4,\N'/4)$.
Our task is now to explicitly evaluate this integral expression.
To do this, we will follow closely the one-instanton
calculation for $\N=2$ SQCD performed in Sec.~8 of Ref.~\cite{KMS}.

The first step is to exponentiate the $\delta$-function
constraints in the measure $\int d{\tilde{\mu}}^{\1I}_{\sst SQCD}$ by
introducing a set of auxiliary (Lagrange multiplier) integration
variables. Thus, for the ``spin-$1$'' constraints on $a$,
we write
\be
\prod_{c=1}^{3} \delta\left(\hf(\tau^c)^\dalpha_{\ \dbeta}\,
\wbar_{\ u}^\dbeta w_{u\dalpha}\right) = \frac{1}{\pi^3} \int d^3p\,
\exp\left(ip^c (\tau^c)^\dalpha_{\ \dbeta} \,\wbar_{\ u}^\dbeta w_{u\dalpha}
\right),
\ee
where the $p^c$ are a triplet of standard bosonic Lagrange multipliers,
and for the ``spin-$1/2$'' constraints on $\M$ and $\N$, we write
\ba
\prod_{\dalpha=1,2} \delta\left(\mubar_u w_{u\dalpha} + \wbar_{\dalpha u}
\mu_u\right)
&=& 2\int d^2\xi \,\exp\left(\xi^\dalpha(\mubar_u w_{u\dalpha}
+\wbar_{\dalpha u}\mu_u)\right),\nn\\
\prod_{\dalpha=1,2} \delta\left(\nubar_u w_{u\dalpha} + \wbar_{\dalpha u}
\nu_u\right)
&=& 2\int d^2\eta \,\exp\left(\eta^\dalpha(\nubar_u w_{u\dalpha}
+\wbar_{\dalpha u}\nu_u)\right),
\ea
where the Grassmann spinors $\xi^\dalpha$ and $\eta^\dalpha$ serve
as fermionic Lagrange multipliers.
The exponentiation of the ``spin-$0$'' $\delta$-function
constraint on $\Atot$ is accomplished in a more subtle way
involving the $\Atot$-dependent terms in the instanton action.
We write
\ba
&&\int
d\Atot \, \delta\left(\bigL\cdot\Atot-(\Lambda+\Lambda_f)\right)
\exp\left(8\pi^2(\Lambdabar+\Lambdahyp+\Lambda^{\prime}_{\hyp})
\Atot\right)\nn\\
&&\equiv {1\over\bigL}
\exp\left(8\pi^2(\Lambdabar+\Lambdahyp+\Lambda^{\prime}_{\hyp})
\cdot\bigL^{-1}\cdot(\Lambda+\Lambda_f)\right)\nn\\
&&= 8\pi\int d(\Re z)\,d(\Im z) \,\exp\left(-8\pi^2(\zbar\bigL z-
(\Lambdabar+\Lambdahyp+\Lambda^{\prime}_{\hyp})z
-\zbar(\Lambda+\Lambda_f))\right),
\ea
where we have introduced a complex auxiliary integration variable $z$. 

After the exponentiation of the $\delta$-function constraints, the next step
is to integrate over the collective coordinates
$\{w,\mu,\nu,\K,\tilde\K,\R,\tilde\R\}$. This is a fairly
straightforward procedure since all the relevant integrals are Gaussian;
the result is 
\ba
\Foneinst &=&\frac{iC^{\prime}_{1}}{2\pi^2}
e^{-(8\pi^2/g^2)}\int d^3p \,d^2\xi\, d^2\eta\,
d(\Re z)\, d(\Im z) \,\left[ \prod_{u=1}^{N}
\frac{32\pi^6\alphabar_u^2}{(8\pi^2|\alpha_u|^2)^2+
\sum_{c=1,2,3}(p^c+\Xi_{u}^{c})^2}\right]\nn\\
&&\times \left[ \prod_{f=1}^{N_F}(m_f+i\sqrt{2}z) \right] F(z),
\ea
where $\alpha_u = \v_u + iz$ and
$\Xi_{u}^{c} = (\xi_\dalpha (\tau^c)_{\ \dbeta}^{\dalpha} \eta^{\dbeta})
/2\sqrt{2}\pi^2\alphabar_u$
and the factor $F(z)$, which is induced by the symmetric tensor hypermultiplet,
is given by
\ba
F(z)&=& \int d\mu_{\hyp}^{\1I}\,
\exp\left(4\sqrt{2}\pi^2\tilde{\rho}_u\v_u\rho_u
+8\pi^2\Lambda_{\hyp}^{\prime}z
-2m\pi^2(2\tilde{\rho}_u\rho_u
+\tilde{\R}^{\prime\alpha}\R^{\prime}_{\alpha}) \right)\nn\\
&=& M^{-(N+2)}
\left[ \prod_{u=1}^{N} \left(m+\sqrt{2}iz-\sqrt{2}v_u\right)\right]
\left(m+2\sqrt{2}iz\right)^2.\label{hypfactor}
\ea

It remains to integrate over the auxiliary integration
variables $\{p,\xi,\eta,z\}$. Since the hypermultiplet-induced
factor $F(z)$ has no dependence on $p$, $\xi$ or $\eta$,
the integration over these variables is exactly as described 
in Sec.~8 of Ref.~\cite{KMS}. Thus we arrive at the expression
\ba
\Foneinst &=& \frac{i\Lambda^{b_0}_{PV}}{2^N\pi^2}
\left(\sum_{u=1}^N{\partial\over\partial\vbar_u}\right)^2\cdot
\int d(\Re z)\,d(\Im z)\,\left\{\sum_{u=1}^N
{\alphabar_u\over\alpha_u}\left[\prod_{v\neq u}
{\alphabar^2_v\over|\alpha_v|^4-|\alpha_u|^4}\right]\right.\nn\\
&&\left.\times\left[\prod_{f=1}^{N_f}(m_f+\sqrtwo iz)\right]
\left[ \prod_{w=1}^{N} (m -\sqrtwo\v_w + \sqrtwo iz) \right]
\left(m + 2\sqrtwo iz\right)^2\right\}.
\ea
(Here we have introduced the dynamical scale $\Lambda_{PV}$, defined by
$\Lambda_{PV}^{b_0}=M^{b_0}\exp(-8\pi^2/g^2)$, where $b_0=N-2-N_F$ is the
first $\beta$-function coefficient.)
As in~\cite{KMS}, we evaluate this final integral by rewriting
the operator $(\sum_u \partial/\partial\vbar_u )^2$ as
$-\partial^2/\partial \bar{z}^2$ acting inside the integral and then
working out the residues of the various poles. The result is
\ba\label{F1final}
&&\!\!\!\!\!\! \Foneinst
= \frac{\Lambda^{b_0}_{PV}}{2^{N}\pi i}
\left\{\sum_{u=1}^N\sum_{v\neq u}
{1\over({\rm v}_v-{\rm v}_u)^2}
\left[\prod_{w\neq u,v}{1\over({\rm v}_w-{\rm v}_u)
({\rm v}_w-{\rm v}_v)}\right]
\left[\prod_{f=1}^{N_f}\left( m_f-\frac{1}{\sqrtwo}(\v_u+\v_v) \right)\right]
\right.\nn\\
&&\left.\qquad\times
\left[\prod_{w=1}^{N}\left(m-\sqrtwo{\rm v}_w
-\frac{1}{\sqrtwo}({\rm v}_u+{\rm v}_v)\right)\right]
\left(m - \sqrtwo ({\rm v}_u+{\rm v}_v)\right)^2\right\}.
\ea

Now we compare our first-principles result with the exact solution derived
in~\cite{LLL}. The prediction for $\F_1$
extracted from this solution is~\cite{ENRS}\footnote{We
have accounted for a factor
of 2 difference in the definition
of the prepotential in~\cite{ENRS} and a factor of $-\sqrt{2}$ 
difference in the definition of the hypermultiplet masses.
We have also eliminated the dynamical scale $\Lambda^{b_0}$ 
appearing in~\cite{ENRS} in favor of $\Lambda^{b_0}_{PV}$ by means
of the relation $\Lambda^{b_0} = (-1)^{N_F} 2^{-b_0/2+2}\Lambda_{PV}^{b_0}$,
which follows from renormalization group matching arguments~\cite{FP}.}
\ba\label{F1curve}
\Foneinst &=& \frac{\Lambda^{b_0}_{PV}}{2^{N}\pi i} 
\left\{\sum_{u=1}^N \left[\prod_{v\neq u}
{1\over({\rm v}_v-{\rm v}_u)^2}\right]
\left[\prod_{f=1}^{N_f}\left(m_f-\sqrtwo\v_u\right)\right]\right.\nn\\
&&\left.\times
\left[\prod_{w=1}^{N}\left(m-\sqrtwo({\rm v}_u+{\rm v}_w)
\right)\right]\left(m - 2\sqrtwo{\rm v}_u\right)^2\right\}.
\ea
Whilst this is not identical to the expression~\rf{F1final}
for general $N_F$, it is equivalent to that expression
for $N_F<N-2$, which is the range of validity of the curve
presented in~\cite{LLL}.
One can show this by checking that the singularity structure
of the two expressions is the same at the points $\v_u-\v_v=0$.
Hence the difference between the two expressions
must be given by a regular function of the VEVs; by dimensional
analysis, this function must be zero for $N_F<N-4$, equal to a numerical
constant in the case $N_F=N-4$, and given by a linear combination of the
masses, $A m + B\sum_f m_f$, where $A$ and $B$ are numerical constants,
in the case $N_F=N-3$. (In this last case, there can
be no linear dependence on the VEVs since $\sum_u \v_u\equiv 0$.)
Since information about the low-energy physics is obtained by
differentiating the prepotential at least once with respect to the
VEVs, it follows that the two expressions are indeed physically
equivalent for $N_F<N-2$. Thus our first-principles one-instanton
calculation has provided a successful test of the exact solution and the
M-theory method that was used to derive it.

It would be desirable to perform instanton tests of the
exact results for $SU(N)$ models with matter hypermultiplets in other
representations such as the antisymmetric and the adjoint.
The instanton calculation is more of a challenge here
since the collective coordinates associated with the hypermultiplets
are not unconstrained (even at the one-instanton level) and one must
introduce extra Lagrange multipliers to deal with the extra
$\delta$-functions in the integration measure.
We aim to address this challenge elsewhere.

\bigskip
\centerline{{\tiny ****************************}}
\medskip
\noindent I would like to thank Valya Khoze for helpful comments
on a draft of this paper.

\end{document}